\documentclass[11pt,english]{article}
\usepackage[T1]{fontenc}
\usepackage[koi8-r]{inputenc}
\usepackage{amsmath}
\usepackage{amssymb}
\usepackage{esint}

\makeatletter
\@ifundefined{date}{}{\date{}}
\AtBeginDocument{\DeclareFontEncoding{T2A}{}{}}\AtBeginDocument{\DeclareFontEncoding{T2A}{}{}}\AtBeginDocument{\DeclareFontEncoding{T2A}{}{}}\AtBeginDocument{\DeclareFontEncoding{T2A}{}{}}%%%%%%%%%%%%%%%%%%%%%%%%%%%%%%%%%%%%%%%%%%%%%%%%%%%%%%%%%%%%%%%%%%%%%%%%%%%%%%%%%%%%%%%%%%%%%%%%%%%%%%%%%%%%%%%%%%%%%%%%%%%

\newtheorem{theorem}{Theorem}
\newtheorem{corollary}{Corollary}
\newtheorem{lemma}{Lemma}

\newtheorem{remark}{Remark}

\usepackage{babel}

\makeatother

\usepackage{babel}
\begin{document}

\title{Convergence to equilibrium for many particle systems}

\author{Lykov A. A., Malyshev V. A.\thanks{Lomonosov Moscow State University, Faculty of Mechanics and Mathematics,
Vorobyevy Gory 1, Moscow, 119991, Russia}}
\maketitle
\begin{abstract}
The goal of this paper is to give a short review of recent results
of the authors concerning classical Hamiltonian many particle systems.
We hope that these results support the new possible formulation of
Boltzmann's ergodicity hypothesis which sounds as follows. For almost
all potentials, the minimal contact with external world, through only
one particle of $N$, is sufficient for ergodicity. But only if this
contact has no memory. Also new results for quantum case are presented.

Keywords: Markov processes, Boltzmann hypothesis, quantum controllability 
\end{abstract}
\tableofcontents{}

\part{Introduction }

The goal of this paper is to give a short review of recent results
(\cite{LM_1})-(\cite{LM_6}) of the authors. In these results classical
Hamiltonian many particle systems were considerd from new point of
view: they have minimal contact with external world, for example only
one particle of $N$ can have this contact. Thus, we consider Hamiltonian
systems with minimal possible randomness. In despite of this, ergodicity
can be proved for almost all potentials of a wide natural class.

Also we present some new results concerning quantum situation and
discuss common points and difference of our results with other research
in mathematical physics and Markov chains theory, for example with
\cite{Llyod,Jurdjevic,WeaverTimeOptimal,WeaverUniv}, \cite{Altafini,B-J-M_2014},
\cite{Kifer,MT,Orey,piece-wise-det,Revuz,Skorohod}.

\subsection{Intro to intro}

To start with, we give very simple intuition. Let finite set $X$
be given, and let $M(X)$ be the set of probability measures on $X$.
Stochastic matrix $P$ defines linear map $P:M(X)\to M(X)$, and discrete
time Markov chai\b{n} $\xi_{k}$ depending on the initial distribution
$\mu_{0}$ of $\xi_{0}$.

It is known that if the matrix $P^{k}$, for some $k$, has all elements
positive, then there exists unique $P$-invariant measure $\pi$,
and moreover for any initial measure $\mu_{0}$ as $n\to\infty$ 
\begin{equation}
P^{n}\mu_{0}\to\pi,\label{discrete_strong_erg}
\end{equation}
that is for any $A\subset X$ the sequence of real numbers $(P^{n}\mu_{0})(A)$
converges to $\pi(A)$. We will call this property strong ergodicity.
A weaker ergodicity property (we will call it Cesaro ergodicity) 
\begin{equation}
\frac{1}{n}\sum_{k=1}^{n}P^{k}\mu\to\pi\label{discrete_Cesaro_erg_1}
\end{equation}
follows. It can be formulated differently: for any real function $f(x)$
on $X$ and for any initial state $\xi_{0}$, time averages are approximately
equal to space averages (this was Boltzmann's formulation of his famous
hypothesis in statistical physics). Exact formulation for this could
be: 
\begin{equation}
\frac{1}{n}\sum_{k=1}^{n}f(\xi_{k})\to\sum_{x\in X}f(x)\pi(x)\label{discrete_Cesaro_erg_2}
\end{equation}
as $n\to\infty$, with $L_{1}$-convergence, or in some other sense.

Deterministic map $U:X\to X$ is a particular case of Markov chains
- when any element of matrix $P$ is either $0$ or $1$. We will
consider only one-to-one maps $U$. Then it is clear that 
\begin{enumerate}
\item if $N=|X|>1$ then strong convergence never holds because any $U$
defines a partition $X=X_{1}\cup...\cup X_{m}$ such that $U$ is
cyclic on any $X_{i}$; 
\item convergence of 
\[
\frac{1}{n}\sum_{k=1}^{n}f(U^{k}x)
\]
holds for any $U$ and any $x$ (this is a trivial case of the famous
Birkhoff-Khinchin ergodicity theorem) but Cesaro ergodicity (the limit
is the unique invariant measure) holds iff there is only one cycle; 
\item note that there are $N^{N}$ deterministic maps, among them $N!$
one-to-one maps, and among the latter only $(N-1)!$ maps with unique
cycle. Thus, Cesaro ergodicity is also a rare event but not so rare
as strong ergodicity. 
\end{enumerate}
If the set $X$ is not finite, for example a smooth manifold, the
situation becomes enormously more complicated. Ludwig Boltzmann did
not give exact mathematical formulations. Later on, various formulations
of the problem appeared. For some history of ergodicity theory we
refer to \cite{Szasz} and references therein. What could be the ways
to avoid extreme complexity ? First of all, one must find wider and
possibly alternative exact formulations of the problem.

\subsection{Classical ergodicity}

We will consider $N$-particle systems with arbitrary but finite $N$.
Namely, $N$ point particles in (coordinate) space $R^{d}$ with coordinate
vectors $q_{i}=(q_{i1},...,q_{id})$, velocities $v_{i}=(v_{i1},...,v_{id})$,
momenta $p_{i}=m_{i}v_{i}$ and with the interaction defined by smooth
symmetric potentials $V_{ij}(q_{i},q_{j})$. To avoid double indices
we write further $N$ instead of $dN$ (thus the index $i$ should
be considered as a pair (particle number, coordinate number)). Then
the dynamics in the phase space $R^{2N}=\{\psi=(q_{1},...,q_{N},p_{1},...,p_{N})\}$
is defined by the following system of Hamiltonian equations 
\begin{equation}
\frac{dq_{k}}{dt}=\frac{\partial H}{\partial p_{k}}=v_{k},\,\,\,\frac{dp_{k}}{dt}=-\frac{\partial H}{\partial q_{k}},k=1,...,N,\label{closed_equations}
\end{equation}
with the Hamiltonian 
\[
H=H(\psi)=\sum_{k=1}^{N}\frac{p_{k}^{2}}{2m_{k}}+Q,\,\,Q=\sum_{1\leq k\leq l\leq N}V_{kl}(q_{k},q_{l}).
\]
This dynamics defines a one parameter group of one-to-one transformations
$U^{t}:R^{2N}\to R^{2N}$ of the phase space. We assume that $Q\to+\infty$
as $\max_{k}|q_{k}|\to\infty$, so that no particle could escape to
infinity. This assumption is similar to assuming the system to be
in some finite volume (system in the box) $\Lambda$ with reflecting
boundary $\partial\Lambda$. Then the energy surface $M_{h}=\{\psi:H(\psi)=h\}\subset R^{2N}$
is bounded for any $h$ and (by the energy conservation law) is invariant
with respect to this dynamics.

Liouville's theorem says that on any $M_{h}$ there exists finite
probability measure $\lambda_{h}$ (Liouville's measure - normalized
restriction of the Lebesgue measure $\lambda$ on the phase space),
invariant with respect to this dynamics.

We say that for a given $H$ the system is ergodic if for any $\psi\in M_{h}$
\begin{equation}
\lim_{T\to\infty}\frac{1}{T}\int_{0}^{T}f(U^{t}\psi)dt=\int_{M_{h}}f(\psi)d\lambda_{h}(\psi)\label{ergodicity}
\end{equation}
in some space of measurable functions. It can be $L_{2}$, $L_{1}$-convergence,
or uniform convergence with respect to $\psi$.

Possible problems could be the following:

1. Give examples of $V_{kl}$ or, better, classes of $V_{kl}$ such
that ergodicity holds for all sufficiently large $N$. We do not know
any such example, but there are many counterexamples: linear (due
to invariant tori) and non-linear integrable systems;

2. prove the contrary: for typical (or almost any) $Q$, from interesting
classes of potentials, ergodicity does not hold. Of course, one should
define what means <\textcompwordmark{}<almost all>\textcompwordmark{}>;

3. instead of problems 1 and 2 one could look for more natural problems.
What are the reasons for this:

a) the $N$-particle systems discussed above are closed systems, but
it is not known whether closed systems exist in nature. More realistic
is to assume that any system always has some contact with external
world. Then the first natural question is: how weak can be this contact
for the system to be ergodic;

b) this does not contradict to the recent development of theoretical
physics, where the notion of space point itself becomes an approximation.
Namely, in quantum physics the dynamics is defined not as a transformation
of the phase space, but of $L_{2}$ space of functions on the coordinate
space. Moreover, in modern physics the notion of space itself is being
reconsidered: discrete space, quantum space or no space at all.

\subsection{Systems with minima\d{l} randomness }

We consider three types of models with minimum randomness, or combinations
of them.

\paragraph{1. Only one degree of freedom open to external influence }

Namely, we change only one equation (for $k=1$) of the system (\ref{closed_equations})
\begin{equation}
\frac{dq_{1}}{dt}=\frac{\partial H}{\partial p_{1}}=p_{1},\,\,\,\frac{dp_{1}}{dt}=-\frac{\partial H}{\partial q_{1}}+F(x_{1},p_{1},t),\label{W_1}
\end{equation}
where we assume unit masses. All other equations (\ref{closed_equations})
are left unchanged. This is used in part III.

\paragraph{2. Two deterministic evolutions with switching at random time moments}

Let $U_{i}^{t},i=1,2,t\in[0,\infty),$ be two semigroups of deterministic
transformations (of some set $X$). For example, when $X$ is the
phase space of $N$-particle system, and the equations are of the
type (\ref{closed_equations}). Consider the sequence 
\begin{equation}
0=t_{0}<t_{1}<t_{2}<...\label{time_sequence}
\end{equation}
of time moments and denote $\tau_{n}=t_{n}-t_{n-1}$. For any integer
$m\geqslant1$ and non negative real $\tau_{1},\tau_{2},\ldots\tau_{2m+1}\geqslant0$
consider the following transformations 
\[
J_{0}=E,\,\,\,J_{1}=U_{1}^{\tau_{1}},
\]

\begin{equation}
J_{2m}(\tau_{1},\ldots,\tau_{2m})=U_{2}^{\tau_{2m}}U_{1}^{\tau_{2m-1}}\ldots U_{2}^{\tau_{2}}U_{1}^{\tau_{1}},\label{J_2m}
\end{equation}

\begin{equation}
J_{2m+1}(\tau_{1},\ldots,\tau_{2m+1})=U_{1}^{\tau_{2m+1}}J_{2m}(\tau_{1},\ldots,\tau_{2m}),\label{J_2m+1}
\end{equation}
and define the evolution $W(t)$: 
\begin{equation}
W(t)=U_{1}^{t-t_{2m}}J_{2m}(\tau_{1},\ldots,\tau_{2m}),t_{2m}\leq t<t_{2m+1},\label{W_2}
\end{equation}

\[
W(t)=U_{2}^{t-t_{2m+1}}J_{2m+1}(\tau_{1},\ldots,\tau_{2m+1}),t_{2m+1}\leq t<t_{2m+2}.
\]
Define also the following sets of transformations 
\[
\mathcal{J}_{n}(U_{1}^{t},U_{2}^{t})=\{J_{n}(\tau_{1},\ldots,\tau_{n}):\ \tau_{1},\tau_{2},\ldots,\tau_{n}\geqslant0\}.
\]
We say that the triple $(U_{1}^{t},U_{2}^{t},x)$ satisfies the covering
(or contrallaility) condition if there exists $n$ such that 
\begin{equation}
\mathcal{J}_{n}(x)=\mathcal{J}_{n}^{x}(U_{1}^{t},U_{2}^{t})=\mathcal{J}_{n}(U_{1}^{t},U_{2}^{t})x=X,\label{J_n_x}
\end{equation}
that is, starting from $x$, $n$ transformations cover all the set
$X$. The triple satisfies strong covering (strong controllability)
condition if $n$ does not depend on $x$.

Below we always assume \textbf{Condition D}: $\tau_{k}$ are independent
identically distributed positive random variables with $E\tau_{1}<\infty$,
having some density $p(s)=p_{\tau}(s)$ with respect to Lebesgue measure,
positive for all $s\geq0$. However, in some cases weaker assumptions
are possible.

This model is used in section II.2.

\paragraph{3. One deterministic evolution with external deterministic intrusion
at random time moments}

Let be given semigroup $U_{1}^{t}$ and fixed transformation $U_{2}$.
We put $J(t)=U_{2}U_{1}^{t}$ and for $t_{n}\leq t<t_{n+1}$ put 
\begin{equation}
W(t)=U_{1}^{t-t_{n}}J(\tau_{n})...J(\tau_{1}).\label{W_3}
\end{equation}
This model is used in section II.1. Note that in case 2 the trajectories
are continuous, and here not.

\part{Convergence and covering (controllability) property}

\section{Classical dynamics with random time velocity flips}

On the phase space 
\begin{equation}
L=L_{2N}=\mathbb{R}^{2N}=\{\psi=(\begin{array}{c}
q\\
p
\end{array}),q=(q_{1},...,q_{N})^{T},p=(p_{1},...,p_{N})^{T}\in\mathbb{R}^{N}\}\label{space_L}
\end{equation}
($T$ denotes transposition, thus $q,p,\psi$ are the column vectors)
we consider quadratic Hamiltonian 
\begin{equation}
H=H(\psi)=\frac{1}{2}\sum_{i=1}^{N}p_{i}^{2}+\frac{1}{2}\sum_{i,j}V(i,j)q_{i}q_{j}=\frac{1}{2}((\begin{array}{cc}
V & 0\\
0 & E
\end{array})\psi,\psi)_{2}\label{Hamiltonian_quadratic}
\end{equation}
with (symmetric) positive definite matrix $V$, and the corresponding
Hamiltonian system of linear ODE with $k=1,\ldots,N$ 
\begin{equation}
\dot{q}_{k}=p_{k},\dot{p}_{k}=-\sum_{l=1}^{N}V_{kl}q_{l}.\label{Eq_Quadratic}
\end{equation}
Note that here the energy surface $\mathcal{M}_{h}$ is a smooth manifold
(ellipsoid) in $L$ of codimension $1$.

With $(2N\times2N)$-matrix 
\[
A=\left(\begin{array}{cc}
0 & E\\
-V & 0
\end{array}\right)
\]
the system (\ref{Eq_Quadratic}) can be rewritten as 
\begin{equation}
\dot{\psi}=A\psi,\label{LHamVec}
\end{equation}
and the solution of (\ref{LHamVec}) defines the transformation group
\[
U_{1}^{t}=e^{tA}
\]
Now define the transformation $U_{2}$. Assume that at time moments
(\ref{time_sequence}) the following deterministic transformation
$U_{2}:L\to L$ occurs: all $q_{k},p_{k}$ are left unchanged, except
for $p_{1}$, the sign of which becomes inverted 
\[
p_{1}(t_{m}-0)\to p_{1}(t_{m})=-p_{1}(t_{m}-0),m\geq1.
\]
One can say that $U_{2}$ is the velocity flip of the first coordinate
of particle 1. Note that the Liouville measure $\pi$ is invariant
w.r.t. Hamiltonian dynamics and also w.r.t. velocity flips.

For any $t_{n}\leq t<t_{n+1}$ define linear transformations $L\to L$,
putting as in (\ref{W_3}) 
\[
W(t)\psi=e^{(t-t_{n})A}U_{2}e^{\tau_{n}A}...U_{2}e^{\tau_{1}A}\psi,\ \psi\in L.
\]
Thus, we are in the situation of 0.3.3. It is clear that $\mathcal{M}_{h}$
is invariant w.r.t. $W(t)$ for any $h>0$ and $t\geq0$.

\paragraph{What means ``almost all''}

Define the mixing subspace 
\begin{equation}
L_{-}=L_{-}(V)=\{\left(\begin{array}{c}
q\\
p
\end{array}\right)\in L:q,p\in l_{V}\},\label{L_minus}
\end{equation}
where $l_{V}=l_{V,1}$ is the subspace of $\mathbb{R}^{N}$, generated
by the vectors $V^{k}e_{1},\ k=0,1,2\ldots$, where $e_{1},...,e_{N}$
is the standard basis in $\mathbb{R}^{N}$.

Let $\mathbf{V}$ be the set of all positive-definite $(N\times N)$-matrices,
and let $\mathbf{V}^{+}\subset\mathbf{V}$ be the subset of matrices
for which

\begin{equation}
L_{-}(V)=L.\label{L201406141}
\end{equation}
Note that $\mathbf{V}$ can be considered as subset of $R^{\frac{N(N+1)}{2}}$,
thus (the restriction of) Lebesgue measure is defined on it. Let $\omega_{1}^{2},\ldots,\omega_{N}^{2}$
be the eigenvalues of $V$, and let $\mathbf{V}_{ind}$ be the set
of $V\in\mathbf{V}$ such that the square roots $\omega_{1},\ldots,\omega_{N}$
of the eigenvalues are independent over the field of rational numbers.

\begin{lemma}\label{V-sets}

1) The set $\mathbf{V}^{+}$ is open and everywhere dense (assuming
topology of $R^{\frac{N(N+1)}{2}}$) in $\mathbf{V},$

2) The set $\mathbf{V}^{+}\cap\mathbf{V}_{ind}$ is dense both in
$\mathbf{V}^{+}$and in $\mathbf{V}$, and the Lebesgue measures on
$\mathbf{V},\mathbf{V}^{+},\mathbf{V}\cap\mathbf{V}_{ind}$ are all
equal.

\end{lemma}

\paragraph{Covering theorem}

\begin{theorem} \label{th_cover} Assume that $V\in\mathbf{V}^{+}\cap\mathbf{V}_{ind}$,
then there exists $m\geqslant1$ such that for any $\psi\in L$ we
have 
\[
\mathcal{J}_{m}(\psi)=\mathcal{M}_{h}
\]
Moreover, there is the following upper bound on $m$ 
\[
m\leq\frac{2}{\min_{k}\beta_{k}^{2}}+2
\]
where $\beta_{k}=(v_{k},e_{1})$ and $v_{1},...,v_{N}$ are the eigenvectors
of $V$. Moreover, from the properties of $L_{-}$ it follows that
all $\beta_{k}$ are not zero.

\end{theorem}

Similar property was called pure state controllability in quantum
case, see for example \cite{Alessandro,Llyod,WeaverTimeOptimal,WeaverUniv}.

\paragraph{Convergence theorem}

\begin{theorem}\label{th_erg} Assume that $V\in\mathbf{V}^{+}\cap\mathbf{V}_{ind}$.
Then, under condition D, for any initial $\psi(0)$ and any bounded
measurable real function $f$ on $\mathcal{M}_{h}$ we have a.s. 
\[
M_{f}(T)=^{def}\frac{1}{T}\int_{0}^{T}f(\psi(t))dt\to_{T\to\infty}\pi(f)=^{def}\int_{\mathcal{M}_{h}}fd\pi
\]

\end{theorem}

If, for example, $\tau_{i}$ have exponential distribution with the
density $\lambda\exp(-\lambda\tau),\ \lambda>0$, then it defines
Markov process $\psi(t)$ with right continuous deterministic trajectories
and random jumps. Such processes are often called piecewise deterministic
Markov processes, see for example \cite{piece-wise-det}. At the same
time, this can be considered as an example from random perturbation
theory, see \cite{Kifer} where the problem of invariant measures
is studied.

\section{Finite quantum dynamics with random time switching}

Here we consider the situation of the section 0.3.\.{2}, and assume
both groups $U_{i}^{t}$ to be unitary evolutions in $C^{N}$. Examples
could be quantum walks on finite lattices.

\subsection{Definitions and Results}

We consider $\mathcal{H}=C^{N},N>1,$ as the Hilbert space with the
scalar product (hermitian form) 
\[
(\psi,\psi')=\sum_{k=1}^{N}\psi_{k}\bar{\psi}'_{k},\quad\psi,\psi'\in\mathcal{H}.
\]
Let $\mathcal{O}$ be the set of all hermitian (self adjoint) operators
on $\mathcal{H}$. Lie algebra structure on $\mathcal{O}$ is introduced
as 
\begin{equation}
\{A,B\}=i[A,B]=i(AB-BA)\in\mathcal{O}.\label{skobki_figurnye}
\end{equation}

Let $U(N)$ be the group of unitary transformations of $\mathcal{H}$.
Consider two its one-parametric subgroups $U_{k}^{t}=e^{-itH_{k}},\ k=1,2,\ t\geqslant0$,
where $H_{1},H_{2}\in\mathcal{O}$. For any integer $m\geqslant1$
and any real $s_{1},s_{2},\ldots s_{2m+1}\geq0$ consider the transformations
(\ref{J_2m}) and (\ref{J_2m+1}) and define the following sets of
unitary matrices 
\[
\mathcal{J}_{n}(H_{1},H_{2})=\{J_{n}(s_{1},\ldots,s_{n}):\ \tau_{1},\tau_{2},\ldots,\tau_{n}\geq0\}.
\]

\paragraph{$U$-controllability}

We say that the pair $(H_{1},H_{2})$ of hermitian operators satisfies
the $U$-controllability condition, if there exists $n$ such that
\[
\mathcal{J}_{n}(H_{1},H_{2})=U(N).
\]

\begin{theorem} \label{UcontrolTheorem} For the pair of hermitian
operators $(H_{1},H_{2})$ the $U$-controllability condition holds
iff the linear span $\mathcal{L}=\mathcal{L}(H_{1},H_{2})$ (over
the field of real numbers) of the operators 
\begin{equation}
H_{1},H_{2},\{H_{1},H_{2}\},\{H_{1},\{H_{1},H_{2}\}\},\{H_{2},\{H_{1},H_{2}\}\},\ldots\label{generatorsEq}
\end{equation}
coincides with $\mathcal{O}$.

\end{theorem}

The proof of this assertion one can find in many sources: see \cite{WeaverUniv},
also theorem 3.2.1, p. 82 in the book \cite{Alessandro}, also many
references in \cite{WeaverTimeOptimal}. Then $\mathcal{L}$ is a
subalgebra of $\mathcal{O}$ with respect to the operation (\ref{skobki_figurnye}),
and $i\mathcal{L}$ is called the dynamical Lie algebra in \cite{Alessandro}.

Denote $\Sigma\subset\mathcal{O}\times\mathcal{O}$ the set of all
pairs of hermitian operators $(H_{1},H_{2})$, for which the $U$-controllability
property holds. For any $H_{1}\in\mathcal{O}$ define the set $\Sigma(H_{1})$
of all $H_{2}\in\mathcal{O}$ such that the pair $(H_{1},H_{2})$
is $U$-controllable. We shall say that the operator $H_{1}$ is \textbf{almost}
\textbf{$\mathbf{U}$-controllable}, if the set $\Sigma(H_{1})$ is
open and everywhere dense in $\mathcal{O}$.

\begin{theorem}{[}almost all theorem{]} \label{almostAllTh} The
following assertions hold:

1) $\Sigma$ is open and everywhere dense in $\mathcal{O}\times\mathcal{O}$.

2) the set of almost $U$-controllable operators is open and everywhere
dense in $\mathcal{O}$.

\end{theorem}

One can find the formulations of theorem \ref{almostAllTh} in the
papers \cite{WeaverUniv}, \cite{Llyod} and \cite{Jurdjevic}. We
provide below formal rigorous proof. But first we want to give more
constructive criteria. Let $\lambda_{1},\ldots,\lambda_{N}$ be the
eigenvalues of $H_{2}$ and $\psi_{1},\ldots,\psi_{N}$ be the corresponding
eigenvectors.

\begin{theorem} \label{controlExTh} Assume that the following two
conditions hold: 
\begin{enumerate}
\item $(H_{1}\psi_{k},\psi_{j})\ne0$ for all $k\neq j$; 
\item $\lambda_{k}-\lambda_{l}\ne\lambda_{k'}-\lambda_{l'}$ for any ordered
pairs $(k,l)\ne(k',l')\in\{1,\ldots,N\}^{2}$. 
\end{enumerate}
Then for $\mathcal{L}=\mathcal{L}(H_{1},H_{2})$ the following assertions
hold: 
\begin{enumerate}
\item if $\mathrm{Tr}(H_{1})=\mathrm{Tr}(H_{2})=0$, then $\mathcal{L}$
coincides with the subalgebra of all hermitian operators with zero
trace; 
\item otherwise $\mathcal{L}$ coincides with $\mathcal{O}$. 
\end{enumerate}
\end{theorem}

\begin{corollary}

If the operator $H$ has all eigenvalues different, then $H$ is almost
$U$-controllable.

\end{corollary}

Note that for the second condition of this theorem to hold it is sufficient
that $\lambda_{1},\ldots,\lambda_{N}$ were linearly independent over
$\mathbb{Z}$. One could deduce theorem \ref{controlExTh} from results
of \cite{Altafini}, but we will give below a direct and simpler proof.

\paragraph{Convergence to Haar measure}

Consider the sequence (\ref{time_sequence}) of time moments. For
$t\geqslant0$ define the following continuous time random process
with values in $U(N)$: if $t_{n-1}\leq t<t_{n}$ then 
\begin{equation}
X(t)=J_{n}(\tau_{1},\tau_{2},\ldots,\tau_{n-1},t-t_{n-1}),\quad X(0)=E,\label{Xdef}
\end{equation}
(that is $X(t)=W(t)$ from \ref{W_2}). Define also the discrete time
<\textcompwordmark{}<embedded>\textcompwordmark{}> process 
\[
X_{n}=X(t_{n})=J_{n}(\tau_{1},\tau_{2},\ldots,\tau_{n-1},\tau_{n}).
\]
Note that $X_{n}$ is a Markov chain with values in $U(N)$, but $X(t)$
in general is not Markov. For any Borel subset $A\subset U(N)$ let
$P_{n}(A)$ be the probability distribution of the random variable
$X_{n}$.

Denote $\pi$ the normed Haar measure on $U(N)$.

\begin{theorem}{[}convergence to Haar measure{]} \label{convTh}
Assume Condition D and that the pair $(H_{1},H_{2})$ satisfies the
$U$-controllability condition. Then the following assertions hold: 
\begin{enumerate}
\item $P_{n}\to\pi$ in variation as $n\rightarrow\infty$ exponentially
fast, that is for some positive constants $c>0,q<1$, all Borel subsets
$A\subset U(N)$ and all $n\geq1$ 
\[
|P_{n}(A)-\pi(A)|\leqslant cq^{n};
\]

\item For any bounded measurable function $f$, defined on $U(N)$, a. s.
\[
\frac{1}{T}\int_{0}^{T}f(X(t))\ dt\rightarrow_{T\rightarrow\infty}\int_{U(N)}f(u)\pi(du).
\]

\end{enumerate}
\end{theorem}

\paragraph{Convergence for quantum states}

Consider the set $\mathcal{S}$ of real valued linear functionals
on the algebra $\mathcal{O}$, such that 
\[
F(E)=1,\quad F(A^{2})\geqslant0,
\]
for any $A\in\mathcal{O}$, where $E$ is the identity operator. Remind
that any functional $F\in\mathcal{S}$ can be written as 
\[
F(A)=\mathrm{Tr}(\rho A)
\]
for some non negative definite operator $\rho$ such that $\mathrm{Tr}(\rho)=1$,
that is 
\[
\rho=\rho^{*},\quad(\rho\psi,\psi)\geqslant0,
\]
for all $\psi\in\mathcal{H}$.

Dynamics on the set $\mathcal{S}$ is defined by the differential
(Schrodinger) equation 
\begin{equation}
\frac{d\rho(t)}{dt}=-i[H,\rho(t)],\label{shrodEq}
\end{equation}
having the solution 
\[
\rho(t)=U(t)\rho(0)U^{*}(t),
\]
where $U(t)=e^{-iHt}$ is unitary and $H$ is hermitian.

We will say that the function $\rho(t),\ t\geqslant0,$ converges
in Cesaro sense as $t\to\infty$ to $\rho\in\mathcal{S}$, if 
\[
\lim_{T\rightarrow\infty}\frac{1}{T}\int_{0}^{T}\mathrm{Tr}(\rho(t)A)\ dt=\mathrm{Tr}(\rho A)
\]
for any $A\in\mathcal{O}$. We shall use the notation $\rho(t)\to^{c}\rho$.

\begin{theorem}{[}pure state convergence{]}\label{shrodEqConvTh}

Assume that all eigenvalues of the matrix $H$ are different and $\rho(0)=P_{\psi}$
is the projector on the unit vector $\psi$, then 
\[
\rho(t)\stackrel{c}{\longrightarrow}\sum_{k=1}^{N}|(\psi,\psi_{k})|^{2}P_{\psi_{k}}
\]
as $t\rightarrow\infty$, where $\psi_{1},\ldots,\psi_{n}$ -are the
eigenvectors of $H$, which form the orthonormal basis on $\mathcal{H}$.

\end{theorem}

From this theorem it follows that Cesaro limit of $\rho(t)$ depends
on $\rho(0)$ (thus there is no ``ergodicity'' in this case). Now
consider the case when the Hamiltonian $H$ is time dependent: 
\begin{equation}
\frac{d\rho(t)}{dt}=-i[H(t),\rho(t)].\label{shrodEq2}
\end{equation}
Namely, for the time sequence (\ref{time_sequence}) define $H(t)$
as follows 
\[
H(t)=\begin{cases}
H_{1}, & t_{2k}\leq t<t_{2k+1}\\
H_{2}, & t_{2k+1}\leq t<t_{2k+2}
\end{cases},
\]
for $k=0,1,\ldots$ and some pair of hermitian operators $H_{1},H_{2}$.

Then one can write down the solution of (\ref{shrodEq2}) in terms
of the process $X(t)$ (see (\ref{Xdef})): 
\begin{equation}
\rho(t)=X(t)\rho(0)X^{*}(t).\label{rhoXeq}
\end{equation}
\begin{theorem}{[}mixed state convergence{]}\label{convMixedStateTh2}
Assume that the conditions of the theorem \ref{convTh} hold. Then
for any $\rho(0)\in\mathcal{S}$ with probability one we have 
\[
\rho(t)\stackrel{c}{\longrightarrow}\frac{1}{N}E,t\rightarrow\infty.
\]
\end{theorem}

\paragraph{Generalizations for a weaker controllability condition}

Let us say that the pair of hermitian operators $(H_{1},H_{2})$ is
pure states controllable (or controllable for short), if there exists
$n$, such that for any $\psi\in\mathcal{H}$ 
\[
\mathcal{J}_{n}^{\psi}(H_{1},H_{2})=\mathcal{H}.
\]
It is obvious that this condition follows from the $U$-controllability
condition. The inverse statement in general is not true. Moreover,
in the book \cite{Alessandro} there is a general criterion of when
the pair $(H_{1},H_{2})$ is controllable in terms of $\mathcal{L}$
(theorem 3.4.7). .

Define the random process $\psi(t)=X(t)\psi$ and the embedded chain
$\psi_{n}=X_{n}\psi$ as $\psi\in\mathcal{H}$. Let $P_{n}(\psi,A)$
denote the probability that $\psi_{n}$ belongs to Borel subset $A\subset\mathcal{H}$.
Denote $\pi^{*}$ the uniform measure on $S=\{\psi\in\mathcal{H}:\ (\psi,\psi)=1\}$.

\begin{theorem}{[}measure convergence for weaker controllability{]}
\label{convThRemarks}

Assume condition D and that for the pair $(H_{1},H_{2})$ the (pure
state) controllability condition holds. Then: 
\begin{enumerate}
\item $P_{n}(\psi,\cdot)$ converges to $\pi^{*}$ in variation as $n\rightarrow\infty$
exponentially fast and uniformly in $\psi\in S$, that is for some
positive constants $c>0$ and $q<1$, for any Borel subsets $A\subset\mathcal{H}$
and all $\psi\in S$ 
\[
|P_{n}(\psi,A)-\pi(A)|\leqslant cq^{n}
\]
for all $n\geqslant1$. 
\item For any bounded measurable function $f$ on $S$ and any initial $\psi(0)\in S$
a.s. 
\[
\frac{1}{T}\int_{0}^{T}f(\psi(t))\ dt\rightarrow_{T\rightarrow\infty}\int_{S}f(\psi)d\pi^{*}(\psi).
\]

\end{enumerate}
\end{theorem}

The proof is exactly the same as the proof of theorem \ref{convTh}.

Theorem on mixed states convergence holds also under controllability
condition.

\begin{theorem}{[}mixed state convergence 2{]} \label{convMixedStateTh2-1}

Assume that the conditions of theorem \ref{convThRemarks} hold. Then
for any $\rho(0)\in\mathcal{S}$ as $t\rightarrow\infty$ a.s. 
\[
\rho(t)\stackrel{c}{\longrightarrow}\frac{1}{N}E.
\]

\end{theorem}

\subsection{Proofs}

\subsubsection{Proof of Theorems \ref{almostAllTh} and \ref{controlExTh} }

We prove first the Theorem \ref{controlExTh}.

Further on, all matrices are considered in the basis $\psi_{1},\ldots,\psi_{N}$.
Note that the set 
\[
\mathcal{T}=\{H\in\mathcal{O}:\ (H\psi_{k},\psi_{k})=0,\,\,\,for\,\,all\,\,\,k=1,\ldots,N\}
\]
is a linear real space of dimension $d=N(N-1)$.

\begin{lemma} $\mathcal{T}$ is a subset of $\mathcal{L}$.

\end{lemma}

Proof. Define the operator $T:\mathcal{O}\rightarrow\mathcal{O}$
by 
\[
T(H)=\{H_{2},H\}.
\]

To prove the lemma it is sufficient to show that the real linear space
generated by the matrices 
\begin{equation}
T(H_{1}),T^{2}(H_{1}),\ldots,T^{d}(H_{1}),\ldots\label{thmatrix}
\end{equation}
coincides with $\mathcal{T}$. For the $(k,j)$-th element of the
matrix $T(H)$ we have 
\[
\left(T(H)\right)_{k,j}=h_{k,j}(\lambda_{k}-\lambda_{j})i,
\]
where $H=(h_{k,j})$. It follows that 
\[
\left(T^{n}(H)\right)_{k,j}=h_{k,j}\left((\lambda_{k}-\lambda_{j})i\right)^{n}.
\]
As all elements of the matrix $H_{1}$ are non zero, then, for all
$n$, the linear dependence of $T(H_{1}),T^{2}(H_{1}),\ldots,T^{n}(H_{1})$
over $\mathbb{C}$ is equivalent to the linear dependence of the matrices
$T_{1},\ldots,T_{n}$ over $\mathbb{C}$, where 
\[
(T_{n})_{k,j}=\left((\lambda_{k}-\lambda_{j})i\right)^{n}.
\]
But this is possible (due to the condition on the eigenvalues of $H_{2}$)
only for $n>\frac{N(N-1)}{2}=\frac{d}{2}$. Lemma is proved.

We will need one more lemma. Denote $E_{k,j}\in\mathcal{O}$ the hermitian
operator with the matrix 
\[
(E_{k,j}\psi_{k'},\psi_{j'})=\begin{cases}
1, & k'=j'=k,\\
-1, & k'=j'=j,\\
0, & otherwise.
\end{cases}.
\]
\begin{lemma} \label{Ekjlem} For all $k\ne j=1,\ldots,N$ 
\[
E_{k,j}\in\mathcal{L}.
\]
\end{lemma}

Proof. By symmetry it is sufficient to prove that $E_{1,N}\in\mathcal{L}$.
For this we shall define the operator $S\in\mathcal{T}$, such that
$\{H_{1},S\}=E_{1,N}$. Lemma will follow from this. For any $\psi\in\mathcal{H}$
and $S\in\mathcal{O}$ we have: 
\[
\left(\{H_{1},S\}\psi,\psi\right)=i((S\psi,H_{1}\psi)-(H_{1}\psi,S\psi))=2\mathrm{Im}((H_{1}\psi,S\psi)).
\]
Denote $H_{1}=(h_{k,j})$. Define now the operator $S\in\mathcal{T}$
as follows: 
\begin{align*}
S\psi_{1}= & \ b_{1}\psi_{2},\\
S\psi_{k}= & \ a_{k}\psi_{k-1}+b_{k}\psi_{k+1},\ k=2,\ldots,N-1,\\
S\psi_{N}= & \ a_{N}\psi_{N-1},
\end{align*}
where $b_{k}=\bar{a}_{k+1},\ k=1,\ldots,N-1$ and 
\[
a_{k}=\frac{i}{2h_{k,k-1}},\quad k=2,\ldots,N.
\]

We have: 
\begin{align*}
\left(\{H_{1},S\}\psi_{1},\psi_{1}\right)= & \ 2\mathrm{Im}(h_{2,1}\bar{b}_{1})=1,\\
\left(\{H_{1},S\}\psi_{k},\psi_{k}\right)= & \ 2\mathrm{Im}\left(h_{k-1,k}\bar{a}_{k}+h_{k+1,k}\bar{b}_{k}\right)=\mathrm{Im}\left(-h_{k-1,k}\frac{i}{\bar{h}_{k,k-1}}+h_{k+1,k}\frac{i}{h_{k+1,k}}\right)=0,\\
\left(\{H_{1},S\}\psi_{N},\psi_{N}\right)= & \ 2\mathrm{Im}(h_{N-1,N}\bar{a}_{N})=-1,
\end{align*}
where $k=2,\ldots,N-1$. It follows that for some $T\in\mathcal{T}$
one can write $\{H_{1},S\}=E_{1,N}+T\in\mathcal{L}$. This proves
the Lemma.

Return now to the proof of the theorem. In the first case, $\mathrm{Tr}(H_{1})=\mathrm{Tr}(H_{2})=0$,
it follows from the definition of $\mathcal{L}$ that the trace of
any operator $H\in\mathcal{L}$ is zero. In other words, $i\mathcal{L}\subset su(N)$,
where $su(N)$ is the Lie algebra of the group of special unitary
matrices. The dimension of $su(N)$ is $N^{2}-1$. Moreover, as $E_{1,k}\in\mathcal{L},\ k=1,\ldots,N-1$
are linearly independent and do not belong to $\mathcal{T}$, the
dimension of $\mathcal{L}$ also equals $N(N-1)+N-1=N^{2}-1$. Thus,
$i\mathcal{L}=su(N)$.

Consider now the second case of the Theorem, that is when for some
$k=1,2$ the trace of $H_{k}$ is not zero. Let $D$ be the operator
with diagonal matrix, having the $(j,j)$-th element equal to $(H_{k})_{j,j}$.
As $\mathcal{T}\subset\mathcal{L}$, then $D\in\mathcal{L}$. As the
trace of $D$ is not zero, then $D,E_{1,2},\ldots,E_{N-1,N}$ are
linearly independent. Then, by the arguments similar to those in the
first case above, we get the proof of the Theorem.

Now we will give the \textbf{proof of theorem \ref{almostAllTh}}.

Proof of Theorem

\paragraph{Proof of assertion 1) }

To prove the first statement of the theorem it is sufficient to prove
the following two assertions:

a) $\Sigma$ is the complement to some algebraic set in $\mathcal{O}\times\mathcal{O}$,
that is to the set of zeroes of some system of polynomial equations.

b) the set $\Sigma$ is not empty. This follows from Theorem \ref{controlExTh}.

Let us prove the assertion a). Let $\mathcal{P}$ be the countable
set of all operators (\ref{generatorsEq}). That is the algebra $\mathcal{L}$
is, by definition, the linear span of the vectors from $\mathcal{P}$.
The set $\mathcal{O}$ of all hermitian operators can be considered
as $N^{2}$-dimensional linear space over reals. Then for the set
$S_{1},\ldots,S_{N^{2}}\in\mathcal{O}$ of such operators denote $F(S_{1},\ldots,S_{N^{2}})$
the determinant of the matrix, with rows of which are these vectors.
Note that if $S_{1},\ldots,S_{N^{2}}\in\mathcal{P}$, then 
\[
G_{S_{1},\ldots,S_{N^{2}}}(H_{1},H_{2})=F(S_{1},\ldots,S_{N^{2}})
\]
is the polynomial of the elements of the matrices $H_{1},H_{2}$.
Let $E$ be the complement to $\Sigma$ in $\mathcal{O}\times\mathcal{O}$.
It is clear that $(H_{1},H_{2})\in E$ iff for any $S_{1},\ldots,S_{N^{2}}\in\mathcal{P}$
will be 
\begin{equation}
G_{S_{1},\ldots,S_{N^{2}}}(H_{1},H_{2})=0.\label{polEq}
\end{equation}
But by Hilbert's Basis theorem there exists finite set of polynomials
of the elements of the pair $(H_{1},H_{2})$, with the same set of
zeroes as the set (\ref{polEq}). Thus the first assertion is proved.

\paragraph{Proof of 2)}

We proved above that $\Sigma$ is the complement to some algebraic
set $E$, that is 
\[
E=\{(H_{1},H_{2}):\ F_{1}(H_{1},H_{2})=\ldots=F_{m}(H_{1},H_{2})=0\}
\]
for some polynomials $F_{1},\ldots,F_{m}$ of the coefficients of
the matrices of operators $H_{1},H_{2}$ in some fixed basis. Consider
the following algebraic set in $\mathcal{O}$: 
\[
E(H_{1})=\{H_{2}:\ F_{1}(H_{1},H_{2})=\ldots=F_{m}(H_{1},H_{2})=0\}.
\]
It is clear that for any $H_{1}\in\mathcal{O}$ the set $\Sigma(H_{1})$
is the complement to $E(H_{1})$ in $\mathcal{O}$. Then $\Sigma(H_{1})$
is either open and everywhere dense or empty. The latter possibility
can occur iff for any $k=1,\ldots,N$ the polynomial (considered as
the function of the matrix $H_{2}$) $f_{k}(H_{2})=F_{k}(H_{1},H_{2})$
is identically zero.

Let us show first that the set of almost $U$-controllable $H_{1}$,
that is for which $\Sigma(H_{1})$ is open and everywhere dense, is
open. Let $H_{1}$ be almost $U$-controllable. Then there exists
$H_{2}$, such that $F_{k}(H_{1},H_{2})\ne0$ for some $k=1,\ldots,m$.
Then for all $H$ in some neighborhood of $H_{1}$ the inequality
$F_{k}(H,H_{2})\ne0$ holds, We get from this that $\Sigma(H)\ne\emptyset$,
then $H$ is also almost $U$-controllable. Now it is sufficient to
prove that the set of almost $U$-controllable operators from $\mathcal{O}$
is dense. Now take $H_{2}$ from theorem \ref{controlExTh}. It is
clear that the set of all $H_{1}$ for which $H_{2}\in\Sigma(H_{1})$
is dense. Our statement follows from this.

Proof of the Corollary.

Let us prove that in some orthonormal basis all non-diagonal elements
of the matrix of $H$ are non zero. Let $\psi_{1},\ldots,\psi_{N}$
be the eigenvectors of $H$. For $t\geqslant0$ consider the orthonormal
basis 
\[
\psi_{k}(t)=e^{itS}\psi_{k},\quad k=1,\ldots,N,
\]
where the hermitian operator $S$ is such that $(S\psi_{k},\psi_{j})\ne0$
for all $k,j=1,\ldots,N$. We have 
\[
(H\psi_{k}(t),\psi_{j}(t))=(H(t)\psi_{k},\psi_{j}),\quad H(t)=e^{-itS}He^{itS},
\]
and 
\[
\frac{dH(0)}{dt}=i[H,S].
\]
It follows that 
\[
\left(\frac{dH(0)}{dt}\psi_{k},\psi_{j}\right)=(S\psi_{k},\psi_{j})i(\lambda_{j}-\lambda_{k}),
\]
where $\lambda_{1},\ldots,\lambda_{N}$ are the eigenvalues of $H$
corresponding to $\psi_{1},\ldots,\psi_{N}$. Then for $t\rightarrow0$
\[
(H\psi_{k}(t),\psi_{j}(t))=\lambda_{k}\delta_{k,j}+(S\psi_{k},\psi_{j})i(\lambda_{j}-\lambda_{k})t+\bar{\bar{o}}(t).
\]
Using the assumptions on $H$ and the choice of $S$ we conclude that
for some small $t$ and all $k\ne j=1,\ldots,N$ the following inequality
holds 
\[
(H\psi_{k}(t),\psi_{j}(t))\ne0.
\]
Now take any hermitian operator $H_{2}$, satisfying the conditions
of Theorem \ref{controlExTh} with eigenvectors $\psi_{1}(t),\ldots,\psi_{N}(t)$.
By this theorem $(H_{1},H_{2})$ is $U$-controllable. In the proof
of theorem \ref{almostAllTh} we got that there is the alternative:
either $\Sigma(H_{1})$ is open and everywhere dense or empty. The
theorem is proved.

Note that the condition that all eigenvalues are different is important.
Because one can show that there is no orthonormal basis in which the
matrix elements of the operator 
\[
H=\left(\begin{array}{cc}
\lambda & 0\\
0 & E
\end{array}\right),\ \lambda\ne1,\lambda>0,
\]
are non zero. Nevertheless, from this one cannot state that this $H$
is not almost $U$-controllable.

\subsubsection{Mixed states convergence: theorem \ref{convMixedStateTh2} }

By equality (\ref{rhoXeq}) and Theorem \ref{convTh} about convergence,
we have (further on $u\in U(N)$) 
\[
\rho(t)\stackrel{c}{\longrightarrow}\rho=\int_{U(N)}u\rho(0)u^{*}\ d\pi(u).
\]
For any $g\in U(N)$ we have 
\[
g\rho g^{*}=\int_{U(N)}(gu)\rho(0)(gu)^{*}\ d\pi(u)=\int_{U(N)}u\rho(0)u^{*}\ d\pi(u).
\]
The last equality follows from the invariance of Haar measure with
respect to left and right multiplication. Then for any $g\in U(N)$
\[
g\rho g^{*}=\rho.
\]
It follows that 
\[
\rho=\frac{1}{N}E.
\]
The proof is finished.

\subsubsection{Theorem \ref{convMixedStateTh2-1} for (pure state) controllability
condition}

One can write the initial state as follows 
\[
\rho(0)=\sum_{k=1}^{N}c_{k}P_{\psi_{k}},\quad\sum_{k=1}^{N}c_{k}=1,\ c_{k}\geqslant0,k=1,\ldots,N,
\]
where $\psi_{1},\ldots,\psi_{N}$ is an orthonormal basis of $\mathcal{H}$.
Then 
\[
\rho(t)=\sum_{k=1}^{N}c_{k}P_{\psi_{k}(t)},
\]
where $\psi_{k}(t)=X(t)\psi_{k}$. By theorem \ref{convThRemarks}
concerning convergence with probability 1 we have: 
\[
\lim_{T\rightarrow\infty}\frac{1}{T}\int_{0}^{T}\mathrm{Tr}(P_{\psi_{k}(t)}A)dt=\int_{S}(A\psi,\psi)d\pi^{*}(\psi)=\mathrm{Tr}(\rho A),
\]
where we put 
\[
\rho=\int_{S}\psi\psi^{*}d\pi^{*}(\psi).
\]
Thus, with probability 1 
\[
\rho(t)\stackrel{c}{\longrightarrow}\rho.
\]
As measure $\pi^{*}$ is invariant with respect to unitary transformations,
then for any unitary matrix $u\in U(N)$ we have: 
\[
u\rho u^{*}=\int_{S}(u\psi)(u\psi)^{*}d\pi^{*}(\psi)=\rho,
\]
and it follows that 
\[
\rho=\frac{1}{N}E.
\]

\subsubsection{Theorem \ref{shrodEqConvTh}: pure state convergence}

Use the expansion of vector $\psi$ in the eigenvectors of $H$: 
\[
\psi=\sum_{k=1}^{N}a_{k}\psi_{k},\quad a_{k}=(\psi,\psi_{k}).
\]
Then 
\[
\rho(t)=P_{\psi(t)},\quad\psi(t)=e^{-iHt}\psi.
\]
Let $\lambda_{1},\ldots,\lambda_{N}$ be eigenvalues of $H$ corresponding
to eigenvectors $\psi_{1},\ldots,\psi_{N}$ correspondingly. Then
\[
\psi(t)=\sum_{k=1}^{N}a_{k}e^{-i\lambda_{k}t}\psi_{k}.
\]
For any hermitian operator $A$ we have: 
\[
\mathrm{Tr}(\rho(t)A)=(A\psi(t),\psi(t))=\sum_{k,j}a_{k}\bar{a}_{j}e^{it(\lambda_{j}-\lambda_{k})}(A\psi_{k},\psi_{j}).
\]
As 
\[
\lim_{T\rightarrow\infty}\frac{1}{T}\int_{0}^{T}e^{it(\lambda_{j}-\lambda_{k})}=\begin{cases}
1, & k=j\\
0, & k\ne j
\end{cases},
\]
the theorem is proved.

\subsubsection{Theorem \ref{convTh}}

Here it is convenient to denote $U(N)=\mathcal{M}$.

\paragraph{Convergence for embedded chain}

It is necessary to do some remarks concerning possible proofs. This
assertion could be examined using the general theory of Markov chains
with general state space, see for example \cite{MT,Skorohod,Orey},
as it was done in simpler cases for random walks on groups (see for
example \cite{Grenander} p. 83, theorem 3.2.6). However, our proof
will be based on theorem 4.1 in \cite{LM_6}.

Note that $X_{n}$ is a Markov process, which is not time homogeneous.
But $\xi_{n}=X_{2n}$ will already be time homogeneous Markov process.
We shall study ergodic properties of $\xi_{n}$ and will understand
how they could be related to ergodic properties of $X_{n}$.

Otherwise speaking $\xi_{n}$ on $\mathcal{M}$ can be defined as
follows: 
\[
\xi_{n}=U_{2}^{\tau_{2n}}U_{1}^{\tau_{2n-1}}\xi_{n-1}=X_{2n}g,\ n=1,\ldots,\quad\xi_{0}=g\in\mathcal{M}.
\]
For $g\in\mathcal{M}$ and Borel subset $A\subset\mathcal{M}$ let
$P(g,A)$ be the one step transition probability of the chain $\xi_{n}$.
The probability $P_{m}(A)$, defined at the beginning of this section,
and $P(g,A)$ are connected as follows: 
\begin{align*}
P_{2m}(A)= & \ P^{m}(e,A),\\
P_{2m+1}(A)= & \ \int_{\mathcal{M}}P_{1}(du)P^{m}(u,A),
\end{align*}
for all $m\geqslant1$, where $e\in\mathcal{M}$ is the identity transformation,
$P^{m}(\cdot,\cdot)$--- $m$-th degree of the kernal $P(\cdot,\cdot)$.

Using Theorem \ref{convThEmbedded}, we get: 
\[
|P_{2m}(A)-\pi(A)|=|P^{m}(e,A)-\pi(A)|\leqslant cq^{m},
\]

\[
|P_{2m+1}(A)-\pi(A)|=|\int_{\mathcal{M}}P_{1}(du)P^{m}(u,A)-\pi(A)|\leqslant\int_{\mathcal{M}}P_{1}(du)|P^{m}(u,A)-\pi(A)|\leqslant cq^{m}.
\]
Thus we have proved the first assertion of Theorem \ref{convTh}.

\begin{theorem} \label{convThEmbedded}

Assume that the conditions of Theorem \ref{convTh} hold. Then $P^{n}(g,\cdot)$
converges to $\pi$ in variation as $n\rightarrow\infty$ exponentially
fast and uniformly in $g\in\mathcal{M}$, that is for some positive
constants $c>0,q<1$, all Borel subsets $A\subset\mathcal{M}$ and
all $g\in\mathcal{M}$: 
\begin{equation}
|P^{n}(g,A)-\pi(A)|\leqslant cq^{n}\label{convThEmEq}
\end{equation}
for all $n\geqslant1$.

\end{theorem}

To prove this theorem we shall use theorem 4.1 from \cite{LM_6}.
Let us check the conditions of this theorem, namely that $\xi_{n}$
is a weakly Feller process, and that for some $n\geqslant1$ the measure
$P^{n}(g,\cdot)$ is equivalent to Haar measure $\pi$ for any $g\in\mathcal{M}$.

\begin{lemma} (Condition A2 from \cite{LM_6}) The kernel $P(\cdot,\cdot)$
is a weak Feller, that is for any open $O\subset\mathcal{M}$ the
transition probability $P(g,O)$ is lower semicontinuous in $g\in\mathcal{M}$.

\end{lemma} For any $g$ denote $\mathbf{1}_{g}(s_{1},s_{2})$ the
indicator function on $R_{+}\times R_{+}$, that is $\mathbf{1}_{g}(s_{1},s_{2})=1$
if $J_{2}(s_{1},s_{2})g\in O$, and zero otherwise. Then we have 
\[
P(g,O)=\int_{R_{+}\times R_{+}}\mathbf{1}_{g}(s_{1},s_{2})p(s_{1})p(s_{2})ds_{1}ds_{2}.
\]
Let $g_{n}\rightarrow g,g_{n}\in\mathcal{M}$ as $n\rightarrow\infty$.
Fix $s_{1},s_{2}\geqslant0$ and consider two cases:

1. $J_{2}(s_{1},s_{2})g\in O$, then starting from some $n$ the inclusion
$J_{2}(s_{1},s_{2})g_{n}\in O$ holds, as $O$ is open. That is why
\[
\lim_{n\rightarrow\infty}\mathbf{1}_{g_{n}}(s_{1},s_{2})=\mathbf{1}_{g}(s_{1},s_{2})=1;
\]
2. $J_{2}(s_{1},s_{2})g\notin O$. Then 
\[
\liminf_{n}\mathbf{1}_{g_{n}}(s_{1},s_{2})\geqslant\mathbf{1}_{g}(s_{1},s_{2})=0.
\]
Thus for any $s_{1},s_{2}$ 
\[
\liminf_{n}\mathbf{1}_{g_{n}}(s_{1},s_{2})\geqslant\mathbf{1}_{g}(s_{1},s_{2}).
\]
Then by Fatou lemma 
\[
\liminf_{n}P(g_{n},O)\geqslant P(g,O).
\]
So, the lemma is proved.

\begin{lemma}\label{A1lemma} (Condition A1 from \cite{LM_6}) For
some $m\geqslant1$ the measures $\pi$ and $P^{m}(g,\cdot)$ are
equivalent for any $g$. Moreover, there is exist $m$-step transition
density $p^{m}(g,u)$ measurable on $\mathcal{M}\times\mathcal{M}$
and positive almost everywhere, such that 
\[
P^{m}(g,B)=\int_{B}p^{m}(g,u)d\pi(u)
\]
for all $g\in\mathcal{M}$ and all Borel subset $B\subset\mathcal{M}$.

\end{lemma}

Let us remind that $\xi_{n}$ can be presented as follows 
\[
\xi_{n}=J_{2n}(\tau_{1},\tau_{2},\ldots,\tau_{2n})g,
\]
where operator $J_{2n}$ was defined above. Further on we assume that
$m=2n^{2}$, where $n$ is as in the definition of $U$-controllability.
Introduce the following set 
\[
\Omega_{m}=\{(s_{1},\ldots,s_{m}):\ s_{i}\geqslant0,\ i=1,\ldots,m\}\subset\mathbb{R}_{\geqslant0}^{m}.
\]
For any $g\in\mathcal{M}$ the function $J_{m}^{g}(s_{1},\ldots,s_{m})=J_{m}(s_{1},\ldots,s_{m})g$
acts from $\Omega_{m}$ to $\mathcal{M}$, then by definition of $U$-controllability:
\[
J_{m}^{g}(\Omega_{m})=\mathcal{M}.
\]
\begin{lemma} \label{zeroSet-1} For any measurable $B\subset\mathcal{M}$
its Haar measure $\pi(B)=0$ iff the Lebesgue measure $\lambda$ of
the set $(J_{m}^{g})^{-1}(B)$ in $\Omega_{m}$ is zero.

\end{lemma}

1) Assume that for some $B\subset\mathcal{M}$ we have $\pi(B)=0$.
Let us show that $\lambda((J_{m}^{g})^{-1}(B))=0$. Let $A_{cr}$
be the set of critical points of the map $J_{m}^{g}$ (that is points
$\overline{\tau}=(\tau_{1},\ldots,\tau_{m})$ where the rank of the
Jacobian is not maximal) and let $E=J_{m}^{g}(A_{cr})\subset\mathcal{M}$
be the set of critical values of $J_{m}^{g}$. By Sard's theorem $\pi(E)=0$.
But as $J_{m}^{g}(\Omega_{m})=\mathcal{M}$, then there exists non-critical
point $\overline{\tau}=(\tau_{1},\ldots,\tau_{m})\in\Omega_{m}$,
that is such that the rank of $dJ_{m}^{g}$ at this point equals $N^{2}$.
As the map $J_{m}^{g}$ is analytic in the variables $\tau_{1},\ldots,\tau_{m}$,
the set of points $A_{cr}$, where the rank is less than $N^{2}$,
has Lebesgue measure zero. Then the equality $\lambda((J_{m}^{g})^{-1}(B))=0$
follows from theorem 1 of \cite{Ponomarev}.

2) Assume that for some $B\subset\mathcal{M}$ we have $\pi(B)>0$,
and let us show that $\lambda((J_{m}^{g})^{-1}(B))>0$. By Lebesgue
differentiation theorem there exists point $g'\in\mathcal{M}\setminus E$
and its neighborhood $O(g')$ such that $\pi(O(g')\cap B)>0$. Then
there is point $\overline{\tau}=\overline{\tau}(g')\in(J_{m}^{g})^{-1}(g')$
and some its neighborhood $O(\overline{\tau})\subset\Omega_{m}$,
so that the restriction of $J_{m}^{g}$ on $O(\overline{\tau})$ is
a submersion. Then $\pi(O(g')\cap B)>0$ implies $\lambda((J_{m}^{g})^{-1}(B)\cap O(\overline{\tau}))>0$.
So, lemma \ref{zeroSet-1} is proven.

Denote $p_{\tau}^{(m)}$ the product of $2m$ densities $p_{\tau}$,
then as for any $B\subset\mathcal{M}$ 
\[
P^{m}(g,B)=\int_{(J_{m}^{g})^{-1}(B)}p_{\tau}^{(m)}(\overline{\tau})d\overline{\tau},
\]
by lemma \ref{zeroSet-1} we get that $P^{m}$ and $\pi$ are equivalent
measures.

The proof of measurability of the transition density one can find
in theorem 1, p. 180 of \cite{Skorohod}, and in Proposition 1.1,
p.5, of \cite{Orey}.

So, lemma \ref{A1lemma} is proven.

Let us continue the proof of the assertion of Theorem \ref{convThEmbedded}
concerning convergence of the embedded chain. Let us check that Haar
measure is invariant with respect to $\xi_{n}$. For Borel subset
$B\subset\mathcal{M}$ we have: 
\[
(\pi P)(B)=\int_{\mathcal{M}}d\pi(u)P(u,B)=\int_{\mathcal{M}}d\pi(u)\int_{R_{+}\times R_{+}}\mathbf{1}(U_{2}^{s_{2}}U_{1}^{s_{1}}u\in B)p(s_{1})p(s_{2})ds_{1}ds_{2}=
\]

\[
=\int_{R_{+}\times R_{+}}p(s_{1})p(s_{2})ds_{1}ds_{2}\int_{\mathcal{M}}d\pi(u)\mathbf{1}(U_{2}^{s_{2}}U_{1}^{s_{1}}u\in B)=\pi(B),
\]
where the last equality follows from the invariance of Haar measure
with respect to multiplication.

Further we shall use Theorem 4.1 from \cite{LM_6}. In this paper
there is no assertions concerning geometric rate convergence. However,
during proof of the theorem 4.1 in \cite{LM_6} (see the end of the
proof) the following inequality was proved: 
\begin{equation}
S_{n+k}(A)-I_{n+k}(A)\leqslant(1-\delta)(S_{n}(A)-I_{n}(A)),\ \mbox{for all}\ n=1,2,\ldots,\label{supinfeq}
\end{equation}
where $0<\delta<1$, $k>1$, $A\subset\mathcal{M}$, and 
\[
I_{n}(A)=\inf_{g\in\mathcal{M}}P^{n}(g,A),\quad S_{n}(A)=\sup_{g\in\mathcal{M}}P^{n}(g,A).
\]
But it is obvious that from (\ref{supinfeq}) the assertion (\ref{convThEmEq})
holds. Thus, we have proved the first item of Theorem \ref{convTh}.

\paragraph{Cesaro convergence }

For any measurable bounded function $f$ on $\mathcal{M}$ and any
$T>0$ define the followings integrals 
\[
M_{f}(T)=\frac{1}{T}\int_{0}^{T}f(X(t))\ dt,\quad\pi(f)=\int_{\mathcal{M}}f(u)d\pi(u).
\]

Define the random time $T_{n}$ as 
\[
T_{n}=\sum_{k=1}^{2n}\tau_{k},\quad n=1,2,....
\]

\begin{lemma} For any measurable bounded function $f$ on $\mathcal{M}$
the following limit holds a.s. 
\[
\lim_{n\rightarrow\infty}M_{f}(T_{n})=\pi(f).
\]
\end{lemma}

Proof. Denote $Y_{k}=(\xi_{k},\tau_{2k+1},\tau_{2k+2}),\ k=0,1,\ldots,\ \xi_{0}=e$
the Markov chain with values in $\mathcal{Y}=\mathcal{M}\times\mathbb{R}_{+}\times\mathbb{R}_{+}$.
Then 
\begin{equation}
\int_{T_{k}}^{T_{k+1}}f(X(s))ds=\int_{T_{k}}^{T_{k}+\tau_{2k+1}}f(U_{1}^{s-T_{k}}\xi_{k})ds+\int_{T_{k}+\tau_{2k+1}}^{T_{k+1}}f(U_{2}^{s-(T_{k}+\tau_{2k+1})}U_{1}^{\tau_{2k+1}}\xi_{k})ds=
\end{equation}

\[
=\int_{0}^{\tau_{2k+1}}f(U_{1}^{s}\xi_{k})ds+\int_{0}^{\tau_{2k+2}}f(U_{2}^{s}U_{1}^{\tau_{2k+1}}\xi_{k})ds=F(Y_{k}),
\]
where 
\[
F(g,t_{1},t_{2})=\int_{0}^{t_{1}}f(U_{1}^{s}g)ds+\int_{0}^{t_{2}}f(U_{2}^{s}U_{1}^{t_{1}}g)ds,\ (g,t_{1},t_{2})\in\mathcal{Y}.
\]
Then 
\begin{equation}
M_{f}(T_{n})=\frac{1}{T_{n}}\sum_{k=0}^{n-1}\int_{T_{k}}^{T_{k+1}}f(X(s))ds=\frac{1}{T_{n}}\sum_{k=0}^{n-1}F(Y_{k}).\label{MasSum-1}
\end{equation}
It is easy to show that $Y_{k}$ has invariant measure $\mu=\pi\times P_{\tau},P_{\tau}=p_{\tau}(s_{1})p_{\tau}(s_{2})ds_{1}ds_{2},$
satisfies the conditions of theorem 4.2 in \cite{LM_6} as $\xi_{k}$
satisfies it. Then 
\[
\lim_{n\rightarrow\infty}\frac{1}{n}\sum_{k=0}^{n-1}F(Y_{k})=\mu(F)=\int_{\mathcal{Y}}F(g,t_{1},t_{2})d\mu,
\]
where 
\[
\mu(F)=\int_{\mathbb{R}_{+}\times\mathbb{R}_{+}}P_{\tau}(dt_{1}dt_{2})\int_{\mathcal{M}}d\pi(g)\left(\int_{0}^{t_{1}}f(U_{1}^{s}g)ds+\int_{0}^{t_{2}}f(U_{2}^{s}U_{1}^{t_{1}}g)ds\right)=
\]

\[
=\int_{\mathbb{R}_{+}\times\mathbb{R}_{+}}P_{\tau}(dt_{1}dt_{2})\left(\int_{0}^{t_{1}}ds\int_{\mathcal{M}}d\pi(g)f(U_{1}^{s}g)+\int_{0}^{t_{2}}ds\int_{\mathcal{M}}d\pi(g)f(U_{2}^{s}U_{1}^{t_{1}}g)\right)
\]

\[
=\pi(f)\int_{\mathbb{R}_{+}\times\mathbb{R}_{+}}P_{\tau}(dt_{1}dt_{2})\left(\int_{0}^{t_{1}}ds+\int_{0}^{t_{2}}ds\right)=2\pi(f)E\tau_{1}.
\]
Moreover, by strong law of large numbers for independent random variables
$\tau_{k}$ we have 
\[
\lim_{n\rightarrow\infty}\frac{T_{n}}{n}=2E\tau_{1}.
\]
Then by (\ref{MasSum-1}) we get the proof of the lemma.

To prove the second part of theorem \ref{convTh} we have to estimate
the difference between $M_{f}(T)$ and $M_{f}(T_{n})$. Using the
boundedness $|f(g)|\leqslant c$ we have 
\begin{equation}
|M_{f}(T)-M_{f}(T_{n})|\leqslant|\frac{1}{T}\int_{T_{n}}^{T}f(X(s))ds|+\frac{|T-T_{n}|}{T}|M_{f}(T_{n})|\leqslant\frac{|T-T_{n}|}{T}(c+|M_{f}(T_{n})|).\label{diffEstim-1}
\end{equation}
For any $T>0$ define the random index $n(t)$ so that 
\[
T_{n(t)}\leqslant T<T_{n(T)+1},
\]
and note that 
\[
\frac{|T-T_{n(T)}|}{T}\leqslant\frac{\tau_{2n(T)+1}+\tau_{2n(T)+2}}{T_{n(T)}}=\frac{\tau_{2n(T)+1}+\tau_{2n(T)+2}}{\sum_{k=1}^{2n(T)}\tau_{k}}.
\]
As $E\tau_{1}<\infty$, the law of large numbers, as $n\to\infty$,
gives a.s. 
\[
\frac{\tau_{2N+1}+\tau_{2N+2}}{\sum_{k=1}^{2N}\tau_{k}}\rightarrow0.
\]
But $n(T)\rightarrow\infty$ as $T\rightarrow\infty$. Then the right-hand
side of (\ref{diffEstim-1}) tends to $0$ a.s. as $n=n(T)$ and $T\rightarrow\infty$.
Thus, we complete the proof of theorem \ref{convTh}.

\subsection{From unitary to symplectic }

Here, on a general but very simple example, we show how convergence
in situations with unitary transformations is related to the similar
question for the symplectic transformations. More information one
can find in physical literature, see \cite{buric_1997,heslot_1983,heslot_1985}

We consider $C^{N}$ as complex Hilbert space of dimension $N<\infty$
with the standard basis $e_{n},n=1,2,..,N.$ Then any vector $f\in C^{N}$
can be presented as 
\begin{equation}
f=\sum_{n}\lambda_{n}e_{n},\lambda_{n}=q_{n}+ip_{n}.\label{f_e_n}
\end{equation}
with real $q_{n},p_{n}$. Now let $U^{t}=e^{it\hat{H}}$ be unitary
group in $C^{N}$ where $\hat{H}$ is a selfadjoint operator in $C^{N}$
with matrix 
\[
(a_{kl}+ib_{kl}),a_{kl}=a_{lk},b_{kl}=-b_{lk}.
\]
For the Hamiltonian $\hat{H}$ the quantum dynamics $f(t)=e^{it\hat{H}}f(0)$
for any vector $f(0)\in C^{N}$ satisfies the Schrodinger equation
\begin{equation}
-i\frac{\partial f}{\partial t}=\hat{H}f,\label{schroedinger}
\end{equation}
or

\[
-i\frac{d\lambda_{k}}{dt}=-i(\frac{dq_{k}}{dt}+i\frac{dp_{k}}{dt})=\sum_{l}(a_{kl}+ib_{kl})(q_{l}+ip_{l}),
\]
or 
\begin{equation}
\frac{dp_{k}}{dt}=\sum_{l}(a_{kl}q_{l}-b_{kl}p_{l}),\,\,\,\frac{dq_{k}}{dt}=\sum_{l}(-a_{kl}p_{l}-b_{kl}q_{l}).\label{quantum_equat}
\end{equation}
If we introduce the quadratic Hamiltonian 
\begin{equation}
H=-\frac{1}{2}\sum_{k,l=1}^{N}a_{kl}(q_{k}q_{l}-p_{k}p_{l})+\sum_{k,l=1}^{N}b_{kl}q_{k}p_{l},\label{correct}
\end{equation}
then the equations (\ref{quantum_equat}) coincide with the classical
Hamiltonian equations 
\begin{equation}
\frac{dq_{k}}{dt}=\frac{\partial H}{\partial p_{k}},\frac{dp_{k}}{dt}=-\frac{\partial H}{\partial q_{k}},\label{classical_equa}
\end{equation}
as 
\[
\frac{\partial(\sum_{k,l}b_{kl}q_{k}p_{l})}{\partial p_{k}}=-\frac{\partial(\sum_{k,l}b_{lk}q_{k}p_{l})}{\partial p_{k}}=-\frac{\partial(\sum_{l,k}b_{kl}p_{l}q_{k})}{\partial p_{k}}=-\sum_{k,l}b_{kl}q_{l}.
\]

\begin{remark}

It is interesting that this class of classical Hamiltonian dynamics
has nothing to with the standard Hamiltonian dynamics considered in
section II.1 and Part III. Possible convergence to Liouville and Gibbs
measures of such (gyroscopic) dynamics we shall discuss elsewhere.

\end{remark}

\part{Gibbs equilibrium and memory}

Here we use the notation (\ref{space_L})-(\ref{LHamVec}) from section
II.1, and consider the system (\ref{Eq_Quadratic}) with quadratic
Hamiltonian (\ref{Hamiltonian_quadratic}). Then, the density of Gibbs
measure $\mu_{\beta}$ with respect to Lebesgue measure $\lambda$
on $R^{2N}$ is given by 
\begin{equation}
\frac{d\mu_{\beta}}{d\lambda}=Z_{\beta}^{-1}\exp(-\beta H)=Z_{\beta}^{-1}\exp(-\frac{1}{2}(C_{G,\beta}^{-1}\psi,\psi)_{2}).\label{Gibbs_density}
\end{equation}
So it is gaussian with covariance matrix

\begin{equation}
C_{G,\beta}=\frac{1}{\beta}(\begin{array}{cc}
V^{-1} & 0\\
0 & E
\end{array}).\label{C_G}
\end{equation}
Although Gibbs distribution is invariant with respect to this dynamics,
convergence (for closed system) to it is impossible due to the law
of energy conservation. Thus we have to introduce some random influence,
and we consider the dynamics defined by the system of $2N$ stochastic
differential equations, as in (\ref{W_1}), 
\begin{equation}
\frac{dq_{k}}{dt}=p_{k},\,\,\,\frac{dp_{k}}{dt}=-\sum_{l=1}^{N}V(k,l)q_{l}+\delta_{k,1}(-\alpha p_{k}+f_{t}).\label{system_2N}
\end{equation}
This means that only one degree of freedom, namely $1$ (first coordinate
of the particle $1$) is subjected to damping (defined by the factor
$\alpha>0$) and to the external force $f_{t}$, which we assume to
be a gaussian stationary stochastic process.

\section{Large time behavior for fixed finite $N$}

One can rewrite system (\ref{system_2N}) in the vector notation

\begin{equation}
\frac{d\psi}{dt}=A\psi+F{}_{t},\label{main_matrix_form}
\end{equation}
where 
\begin{equation}
A=(\begin{array}{cc}
0 & E\\
-V & -\alpha D
\end{array}),\label{A}
\end{equation}
$E$ is the unit $(N\times N)$-matrix, $D$ is the diagonal $(N\times N)$-matrix
with all zeroes on the diagonal except $D_{11}=1$, and $F_{t}$ is
the vector $(0,...,0.f_{t},0,...,0)\in R^{2N}$.

\paragraph{Covariance}

All our external forces $f_{t}$ will be gaussian stationary processes
with zero mean. Among them there is the white noise - the generalized
stationary gaussian process having covariance $C_{f}(s)=\sigma^{2}\delta(s)$,
it is sometimes called process with independent values (without memory).
All other stationary gaussian processes, which we consider here, are
processes with memory. We will assume that they have continuous trajectories
and integrable (short memory) covariance 
\[
C_{f}(s)=<f_{t}f_{t+s}>=E(f_{t}f_{t+s}).
\]
Then the solution of (\ref{main_matrix_form}) with arbitrary initial
vector $\psi(0)$ is unique and is equal to 
\begin{equation}
\psi(t)=e^{tA}(\int_{0}^{t}e^{-sA}F{}_{s}ds+\psi(0)).\label{solution_psy_t}
\end{equation}

Our goal, in particular, is to show that even weak memory, in the
generic situation, prevents the limiting invariant measure (which
always exists and unique) from being Gibbs. To formulate more readable
results we assume more: $C_{f}$ belongs to the Schwartz space $S=S(R)$.
Then also the spectral density 
\[
a(\lambda)=\frac{1}{2\pi}\int_{-\infty}^{+\infty}e^{-it\lambda}C_{f}(t)\ dt
\]
belongs to the space $S$.

We shall say that some property (for given $V$) holds for almost
all $C_{f}$ from the space $S$ if the set $S^{(+)}\subset S$ where
this property holds is open and everywhere dense in $S$.

\paragraph{Invariant subspaces}

The subspace $L_{-}\subset L$ was introduced in (\ref{L_minus}).
Now we describe important properties of this set.

\begin{lemma} 
\begin{enumerate}
\item $L_{-}$ and its orthogonal complement denoted by $L_{0},$ are invariant
with respect to the operator $A$. 
\item The spectrum of the restriction $A_{-}$ of $A$ on the subspace $L_{-}$
belongs to the left half-plane, and as $t\to\infty$ 
\[
||e^{tA_{-}}||_{2}\to0
\]
exponentially fast, Moreover\c{ } $L_{-}$ can be defined as 
\[
L_{-}=\{\psi\in L:\ H(e^{tA}\psi)\to0,\ t\to\infty\}\subset L
\]

\end{enumerate}
\end{lemma}

\paragraph{Role of the memory}

\begin{theorem}\label{th_1}

Let $f_{t}$ be either white noise or has continuous trajectories
and integrable $C_{f}$. Then for any Hamiltonian $H$ with $L_{0}=\{0\}$
the following holds: 
\begin{enumerate}
\item there exists gaussian random $(2N)$-vector $\psi(\infty)$ such that
for any initial condition $\psi(0)$ the distribution of $\psi(t)$
converges, as $t\to\infty$, to that of $\psi(\infty)$; 
\item for the process $\psi(t)$ we have $E\psi(t)\to0$ and the covariance
- 
\begin{equation}
C_{\psi(\infty)}(s)=\lim_{t\rightarrow\infty}<\psi(t)\psi^{T}(t+s)>=\lim_{t\rightarrow\infty}C_{\psi}(t,t+s)=W(s)C_{G,1}+C_{G,1}W(-s)^{T},\label{C_psi_t_plus_s}
\end{equation}
where 
\begin{equation}
W(s)=\int_{0}^{+\infty}e^{\tau A}C_{f}(\tau+s)d\tau;\label{W}
\end{equation}

\item For the white noise with variance $\sigma^{2}$ the vector $\psi(\infty)$
has Gibbs distribution (\ref{Gibbs_density}) with the temperature
\[
\beta^{-1}=\frac{\sigma^{2}}{2\alpha};
\]

\item If $\alpha=0,\sigma^{2}>0,$ then for any $i$ the mean energy $EH_{i}$,
where 
\[
H_{i}=\frac{p_{i}^{2}}{2}+\sum_{j}V(i,j)q_{i}q_{j},
\]
of the particle $i$ tends to infinity. If $\alpha>0,\sigma^{2}=0,$
then it tends to zero. 
\end{enumerate}
\end{theorem}

We will use here the shorter notation $C_{\psi(\infty)}(0)=C_{\psi}$.

\begin{theorem}\label{th_2}

Let $N\geq2$, and the Hamiltonian $H$ is such that $L_{0}=L_{0}(H)=\{0\}$.
Then the following assertions hold: 
\begin{enumerate}
\item for any $C_{f}\in S$ the limiting distribution does not have correlations
between coordinates and velocities; 
\item for almost any $C_{f}\in S$ there are non zero correlations between
velocities, that is for some $i\neq j$ $C_{\psi}(p_{i},p_{j})\neq0$.
It follows that the limiting distribution cannot be Gibbs. 
\end{enumerate}
\end{theorem}

\paragraph{Classes of Hamiltonians}

Here we describe classes of potentials with $\dim L_{0}=0$.

Let $\Gamma=\Gamma_{N}$ be connected graph with $N$ vertices $i=1,...,N$,
and not more than one edge per each (unordered) pair of vertices $(i,j)$.
It is assumed that all loops $(i,i)$ are the edges of $\Gamma$.
Denote $\mathbf{H}_{\Gamma}$ the set of (positive definite) $V$
such that $V(i,j)=0$ if $(i,j)$ is not the edge of $\Gamma$. It
is easy to see that the dimension of the set $\mathbf{H}_{\Gamma}$
is equal to the number of edges of $\Gamma$.

Examples can be complete graph with $N$ vertices, or we can consider
the $d$-dimensional integer lattice $Z^{d}$ and the graph $\Gamma=\Gamma(d,\Lambda)$,
the set of vertices of which is the cube 
\[
\Lambda=\Lambda(d,M)=\{(x_{1},...,x_{d})\in Z^{d}:|x_{i}|\leq M,i=1,...,d\}\subset Z^{d}
\]
and the edges $(i,j),\,|i-j|\leq1$.

In general, $V$ is called $\gamma$-local on $\Gamma$ if $V(i,j)=0$
for all pairs $i,j$ having distance $r(i,j)$ between them greater
than $\gamma$, where the distance $r(i,j)$ between two vertices
$i,j$ on a graph is the minimal length (number of edges) of paths
between them.

We shall say that some property holds for almost any Hamiltonian from
the set $\mathbf{H}_{\Gamma}$ if the set $\mathbf{H}_{\Gamma}^{(+)}$,
where the property holds, is open and everywhere dense. Moreover,
the dimension of the set $\mathbf{H}_{\Gamma}^{(-)}=\mathbf{H}_{\Gamma}\setminus\mathbf{H}_{\Gamma}^{(+)}$
where it does not hold, is less than the dimension of $\mathbf{H}_{\Gamma}$
itself.

\begin{lemma}\label{lemma_L_0}

For almost any $H\in\mathbf{H}_{\Gamma}$ we have $\dim L_{0}=0$.

\end{lemma}

\section{Thermodynamic limit}

We have studied above the limit $t\to\infty$ for fixed $N$ and fixed
potential $V=V_{N}$. Here we discuss the limit 
\[
\lim_{N\to\infty}\lim_{t\to\infty}.
\]
First of all, it is not clear that this limit exists, and even less
how it can look like. The only immediate conclusion is that, if it
exists, it will be gaussian, and will depend on the covariance $C_{f}$.
One of the central question is of course: how the limiting distribution
will look like far away from the place of external influence, that
is far away from the particle 1. Will the effect of the memory disappear
or not, that is will this limit have Gibbs covariance or not.

In the white noise case it is easy to prove that we will get anyway
the Gibbs distribution. Consider now the case when $f_{t}$ is not
the white noise. We will prove that for large $N$ the matrices $C_{\psi}$
become close to the simpler matrix 
\[
C_{V}=\frac{\pi}{\alpha}(\begin{array}{cc}
a(\sqrt{V})V^{-1} & 0\\
0 & a(\sqrt{V})
\end{array}),
\]
where $\sqrt{V}$ is the unique positive root of $V$. First of all
note that: 1) $C_{V}$ also defines an invariant measure with respect
to pure (that is with $\alpha=0,f_{t}=0$) Hamiltonian dynamics; 2)
for the white noise case $C_{V}$, corresponds to the Gibbs distribution.

We assume that some graph $\Gamma$ is given with the set of vertices
$\Lambda,|\Lambda|=N$. For any $V\in\mathbf{H}_{\Gamma}$ such that
$L_{0}(V)=\{0\}$, the following representation of the limiting covariance
matrix appears to be crucial 
\[
C_{\psi}=C_{V}+Y_{V},
\]
where $Y_{V}$ is some remainder term.

The following theorem gives the estimates for $Y_{V}$. The norm $||V||_{\infty}$
of a matrix V we define by the formula 
\[
||V||_{\infty}=\max_{i}\sum_{j}|V(i,j)|.
\]

\begin{theorem} \label{th_3} Assume that $V$ is $\gamma$-local
and $||V||_{\infty}<B$ for some $B>0$. Fix also some number $\eta=\eta(N)\geqslant\gamma$.
Then the following assertions hold: 
\begin{enumerate}
\item If $C_{f}\in S$ and has bounded support, that is $C_{f}(t)=0$ if
$|t|>b$ for some $b>0$, then for any pair $i,j$ far away from the
particle 1, that is the distances $r(i,1),\ r(j,1)>\eta(N)$, there
is the following estimate 
\[
|Y_{V}(q_{i},q_{j})|,\ |Y_{V}(p_{i},p_{j})|<K_{0}\left(\frac{K}{\eta}\right)^{\eta\gamma^{-1}}
\]
for some constants $K_{0}=K(C_{f},B,b,\alpha,\gamma)$ and $K=K(C_{f},B,b,\alpha,\gamma)$,
not depending on $N$. 
\item For arbitrary $C_{f}\in S$ the estimate is 
\[
|Y_{V}(q_{i},q_{j})|,\ |Y_{V}(p_{i},p_{j})|<C(k)\eta^{-k},
\]
for any $k>0$ and some constant $C(k)=C(C_{f},k,B,\alpha,\gamma)$,
not depending on $N$. 
\end{enumerate}
\end{theorem}

This theorem allows to do various conclusions concerning the thermodynamic
limit. We give an example.

Fix some $C_{f}(t)\in S$ and some connected countable graph $\Gamma_{\infty}$
with the set of vertices $\Lambda_{\infty}$ and an increasing sequence
of subsets $\Lambda_{1}\subset\Lambda_{2}\subset...\subset\Lambda_{n}\subset...$
such that $\Lambda=\cup\Lambda_{n}$. Let $\Gamma_{n}$ be the subgraph
of $\Gamma_{\infty}$ with the set of vertices $\Lambda_{n}$, that
is $\Gamma_{n}$ inherits all edges between vertices of $\Lambda_{n}$
from $\Gamma$. Here it will be convenient to assume that, for any
fixed $n$, the specified particle (the only one having contact with
external world) has number $N_{n}=|\Lambda_{n}|$. We assume also
that for any $i\in\Lambda_{\infty}$ its distance $r_{n}(i,N_{n})$
to the particle $N_{n}$ tends to $\infty$ as $n\rightarrow\infty$.

Let $l^{\infty}(\Gamma_{\infty})$ be the complex Banach space of
bounded functions on the set of vertices of $\Gamma_{\infty}$: 
\[
l^{\infty}(\Gamma_{\infty})=\{(x_{i})_{i\in\Gamma_{\infty}}:\ \sup_{i\in\Gamma_{\infty}}|x_{i}|<\infty,\ x_{i}\in\mathbb{C}\}.
\]
Fix some $\gamma$-local infinite matrix $V$ on this space and such
that $||V||_{\infty}\leqslant B$. It is clear that $V$ defines a
bounded linear operator on $l^{\infty}(\Gamma_{\infty})$. Denote
$\sigma(V)$ the spectrum of this operator. Let $V_{n}=(V(i,j))_{i,j\in\Lambda_{n}}$
be the restriction of $V$ on $\Lambda_{n}$, it is a matrix of the
order $N_{n}$. Assume that for all $n=1,2,\ldots$ the matrices $V_{n}$
are positive-definite. Note that the condition $L_{-}(V_{n})=L$ may
not hold for some $n$. However, one can choose a sequence of positive-definite
matrices $V'_{n}\in\mathbf{H}_{\Lambda_{n}}$ such that $||V_{n}-V'_{n}||_{\infty}\to0$
as $n\to\infty$ with $L_{0}(V_{n}')=\{0\}$. Moreover, the convergence
of $V'_{n}$ to $V{}_{n}$ can be chosen arbitrary fast. Denote $C_{\psi}^{(n)}$
the limiting covariance matrices corresponding to $V'_{n}$.

\begin{corollary} \label{cor_2} The following assertions hold: 
\begin{enumerate}
\item for any $i,j\in\Lambda_{\infty}$ there exists the thermodynamic limit
\[
\lim_{n\to\infty}C_{\psi}^{(n)}(p_{i},p_{j})=C_{\psi}^{(\infty),p}(i,j),
\]
that is for distribution of velocities; 
\item if for any $i,j\in\Lambda_{\infty}$ there exists finite limits :
\begin{equation}
U(i,j)\doteqdot\lim_{n\rightarrow\infty}V_{n}^{-1}(i,j),\label{0605131}
\end{equation}
then for the coordinates we have 
\[
\lim_{n\to\infty}C_{\psi}^{(n)}(q_{i},q_{j})=C_{\psi}^{(\infty),q}(i,j);
\]

\item assume that the spectral density $a(\sqrt{\lambda})$ is analytic
on the open set containing the spectrum $\sigma(V)$. Then 
\[
C_{\psi}^{(\infty),p}(i,j)=a(\sqrt{V}),
\]
where $a(\sqrt{V})$ is defined in terms of the operator calculus
on $l^{\infty}(\Gamma_{\infty})$ (\cite{Dunford}, p. 568). 
\end{enumerate}
\end{corollary}

In (\cite{LM_4}) one can find all proofs in more general situation
when more than one particle have contact with external world.

\end{document}